\documentclass{article}
\usepackage{graphicx}
\usepackage{amsmath}
\usepackage{amssymb}
\usepackage[numbers]{natbib}
\newcommand{\farcs}{\mbox{\ensuremath{.\!\!^{\prime\prime}}}}
\newcommand{\sun}{\ensuremath{\odot}}

\title{Quasi-Hilda Comet 231P/LINEAR-NEAT: Observation at aphelion using  Himalayan Chandra Telescope (HCT)}

\author{Dhanraj S. Warjurkar \\
              Department of Physics and Astronomy,\\
               Seoul National University, San 56-1, \\
               Sillim-dong, Gwanak-gu, Seoul 151-742, \\
               Republic of Korea.\\
               (Author was affiliated to SNU and now working from India at home) \\
                \texttt{shunya@live.com}           
           \and
           Prasad V. Arlulkar \\
              Hyper Ions Research Labs,\\
               Pune, Maharashtra, 411041, India
}

\begin{document}

\maketitle

\begin{abstract}
Comet 231P/LINEAR-NEAT observed around aphelion  at 4.88 AU, using 2.0 m  Himalayan Chandra Telescope  at Mt. Saraswati, Hanle , India. CCD aperture photometry performed and $R_{C}$-band magnitude $21.37\pm 0.08 $ mag obtained. Comet 231P surface brightness profiles compared with field star and found similar, within  low level noise in the data; its implicate no cometary activity. We measured dust production levels in term of the quantity $Af\rho$, which was estimated 5.0 cm. Most of the comet observed at perihelion but not aphelion, if  we need to calculate total  comet dust contribution in the solar system, this type of study may play important role.   
\end{abstract}

\section{Introduction}
\label{intro}
A Quasi-Hilda Comet (QHC) is a Jupiter family comet (JFC) that interacts strongly with Jupiter and undergoes extended temporary capture by it.  These comets are associated with the Hilda asteroid zone in the 3:2 inner mean-motion resonance with Jupiter. While Di Sisto et al.(2005) suggested that the Hildas and Quasi-Hilda Comets could be closely related to JFCs according to their dynamical behavior\cite{di2005hilda}; Toth give an update  on orbital properties of eliptical comets in Hilda-like orbits and Quasi-Hilda asteroids \cite{Toth2006AA}.   The comet 231P/LINEAR-NEAT  discovered on NEAT Polomar images obtained on 2003 March 10.36 UT and posted on NEO confirmation page, was reported by K Lawence,  a nuclear condensation of diameter about $7\farcs$  and tail about $8 \farcs $ long toward the West. The cometary nature was confirmed by J. Yong at Table Mountain on 2003 March 12.4 UT. The comet 231P is a distant one, with period of 8.08 year.
\par In recent studies shows that short-period comets active between 3 to 7 AU \cite{lowry1999ccd,lowry2001ccd,lowry2003ccd,lowry2005william}. These observation surveys aimed  an investigation of  bared nucleus, but surprisingly high numbers of comets exhibited comae and even dust tails at large ($r_h \geq 3$) heliocentric distance, where volatile sublimation rate is expected to be low,  Chiron is known to be active at solar distance between 8-14 AU \cite{meech1997observations} and was  seen to display substantial out-gassing near aphelion at 17.8 - 18.8 AU between 1969 \& 1977 \cite{bus2001chiron}. According to Meech et al. (2004), the Oort cloud comet C/1987 H1(Shoemakar) exhibited an extensive tail at all distances between 5-18 AU\cite{meech2004comet}. Ohtsuka et al.(2008) studied that Jovion tidal force acting on comet 147P/Kushida-Muramatsu with their QHCs  and perijove passages and inferred that there seems no reason to be belive that  147P was affected by Jovian tides \cite{ohtsuka2008quasi}. Dahlgren \& Lagerkvist (1995), Dahlgren et al. (1997,1999) reported on their photometric survey of the color distribution of the Hilda asteroids \cite{dahlgren1995studyI,dahlgren1997study,dahlgren1999study}. They showed that about 36 \% of the Hildas could be classified as D-type,  28 \% as P-type and only 2\% as C-type. Cheng Y (20013) serendipitous discovery of coma activity of comet 212P which is a Quasi-Hilda object in cometary-like orbit\cite{cheng2013detection}.
\par {Cometary dust particles, trapped in the volatile ices of the nucleus since their formation, contain information on the conditions in the pre-solar nebula. These dust grains are the source of a major component of interplanetary dust and large grains are observed in cometary dust trails forming relatively long lasting meteoroid streams, which gradually dissipate into the zodiacal dust complex. As  Sykes (1992)  pointed out that dust trails are likely to be the principal mass loss mechanism for comets\cite{sykes1992}. Therefore, dust trails may enhance our present understanding of the physical composition of comets, from dirty snowballs to frozen mud balls.  To study the dust production, mass loss rate around aphelion, taxonomical type and correlate this information to other family comets which is not studied, previously. For these reasons, it is important to study dust trails, in order to investigate the said issues. We planned to carry out a comprehensive observation of QHCs at aphelion. Here, we reported photometry result of  comet  231P/LINEAR-NEAT observed at Himalayan Chandra Telescope (HCT). The dust production estimated which is  computed by means of the quantity $Af\rho$. }

\section{Observation}
\label{sec:1}
\begin{figure*}
\centering
  \includegraphics[width=0.75\textwidth]{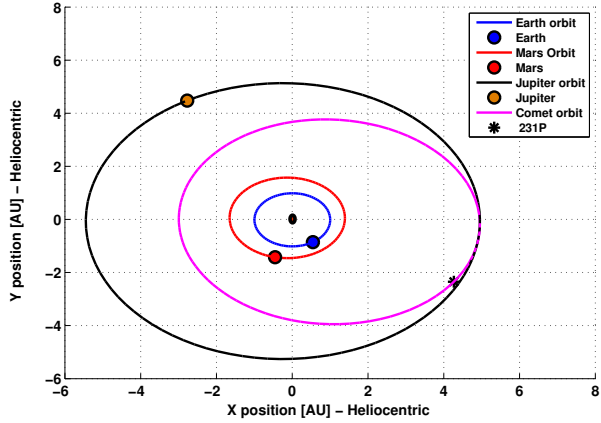}
\caption{Orbit of the comet 231P: Black dot denotes the position of the Sun
and * means the positions of the comet at the time of the observations.}
\label{orp:1}       
\end{figure*}

We used 2.0 m Himalayan Chandra Telescope stands on Mt. Saraswati, Digpa-ratsa Ri, Hanle in south-eastern Ladakh in the eastern Jammu and Kashmir state of India. We observed with the Himalayan Faint Object Spectrograph (HFOSC) at images scale of 0.29 \farcs pixel$^{-1} $ with  2048 $\times$ 4096 pixels CCD camera at the f/9  Cassegrain focus of the telescope through Johnson-Cousins R filter and seeing limited image quality was variable with the range of 0.8 to 1.8 \farcs full width at half maximum (FWHM) with 180 s exposure time and total no. of exposure was 5 ( i.e. 900 seconds). The journal of photometric  observation  shown in the Table \ref{ogc:1} and we observed comet 231P around aphelion at 4.88 AU in order to address the cometary activity issue around aphelion; please, see Fig.\ref{orp:1} for the mean orbital position of comet 231P, as a part of our QHCs observation programme around aphelion, we showed observation position of comet.

\begin{table}
\caption{Observational geometric circumstances}
\label{ogc:1}       
\begin{tabular}{c c c c c c c c}     
\hline
\centering
UT date & RA & DEC & Filter &Exptime &$\alpha$ &$r_h$&$\Delta$\\
& &  &  &  [s]& $[\deg]$ &[AU]  & [AU] \\
\hline \\
2014/07/25.85&22:46:78 & -12:53:57&$R_C$&180 $\times$ 5&7&4.88&4.02\\
\hline
\end{tabular}
\end{table}

\subsection{Data Reduction}
\label{sec:2}
\par All the images were reduced with median images of bias and sky flat using NOAO's Image Reduction and Analysis Facility (IRAF) software package\cite{tody1986iraf}. The pixel-to-pixel sensitivity variation was removed from the data frames using nightly sequences of bias and flats. The variation in sensitivity of the detector was $\pm$ 5\%  from the average. From the comet 231P images,  the instrumental magnitude  obtained using aperture photometry. The flux calibration was done using standard stars in the Landolt catalog 
\cite{landolt1992broadband,landolt2009ubvri} mainly  or field stars listed in the USNO--A2.0 catalogue, occasionally. The comet 231P and stars position were matched with USNO--A2.0~catalogue \cite{monet1998usno}using  WCSTools from NOAO's $IRAF$ software package. For the each of observing session, the diameter of the aperture for the aperture photometry was chosen on the basis of our measurement of the seeing disk size, so that the photometry remains unaffected by trailing \cite{jewitt1987ccd}.

\section{Result}
\label{sec:1}
\subsection{Appearances}
\label{sec:2}
\begin{figure}
   \centering
   \includegraphics[width=0.75\columnwidth ]{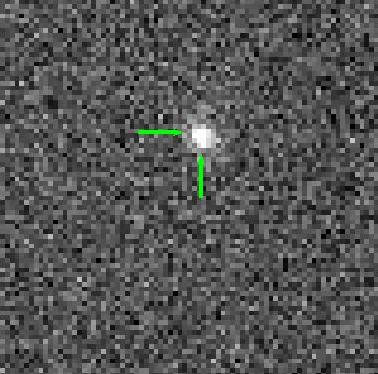}
      \caption{ Image showing comet 231P, made up of $5 \times 180$ s exposures and image scales is $23 \times 23 $ arcseconds in the X and Y directions. The comet appear to inactive; the frames are combined in such a way as to remove background sky, fixed objects and cosmic rays. }
         \label{fitim}
   \end{figure}

\begin{figure*}
\centering
  \includegraphics[width=0.85\textwidth]{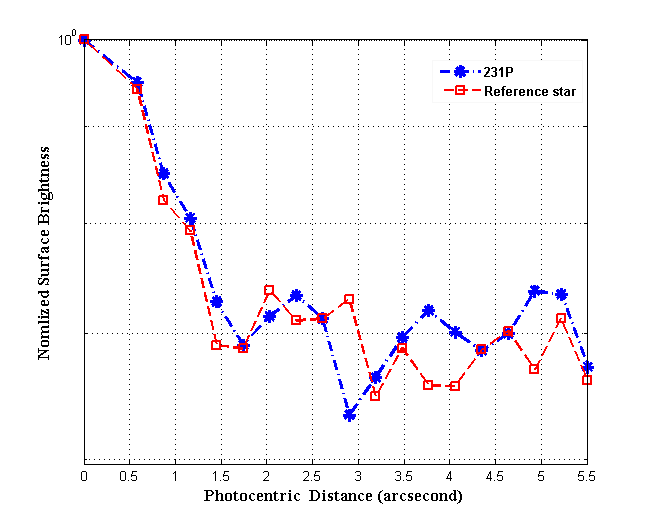}
\caption{Normalized surface brightness profile of 231P, showing surface brightness in
magnitudes per square arcsecond against photocentric distance in arcsecond.}
\label{comet231pcsb2}       
\end{figure*}

We selected R$_\mathrm{C}$-band images taken at HCT (see Fig.\ref{fitim}) and examined a weak coma component existence by comparing the surface brightness-profile of  comet 231P with PSF of stars as explained in the \cite{luu1992near} paper. Both normalized  surface brightness profile were similar, however small differences at the low signal level attributed to noise in the data (see Fig.\ref{comet231pcsb2}).
\subsection{Surface Brightness Profiles and Search of Cometary Activity}
\label{sec:3}
As shows in Fig.\ref{comet231pcsb2} , the surface brightness profiles of the near nuclear coma with respect to the radial distance $\rho $ . The total  R$_\mathrm{C}$-band comet 231P, measured in an aperture equivalent in radius to $\rho = 5.073\times 10^3 $ km at the distance of the comet in given in the table \ref{pr:2}. We created model point source then added come of varying level and finally convolved these in seeing. The coma level were parametrized by \cite{luu1992near,hsieh2005search,kasuga2008obs,ishiguro2011search}
\begin{equation}
\eta = \frac{C_c}{C_n} = \frac{I_c}{I_n} 
\label{cnration}
\end{equation}
where $C_c$ and $C_n$ are the scattering cross-section of coma and nucleus, respectively, which is corresponds to the ratio of the flux density scattered by the coma $I_c$ to the flux density scattered by the nucleus cross-section $I_n$. This is able to characterize varying nucleus and coma contribution levels on preconvolution models.  The parameter $\eta$ can take on the value $\eta \geq$, where $\eta =0$ denotes no comet activity, here we found $\eta=0.01$. The comparison of said mode to comet 231P profile produces no indication of cometary activity within the detection limit.

\subsection{Comet Absolute Magnitude and $Af\rho$ parameter}
\label{sec:4}
The absolute magnitude of comet 231P derived which correspond to the magnitude at an hypothetical point in the space at solar phase angle of $\alpha=0 \deg $, heliocentric distance of  $r_h=1$ AU and an observer  distance  of $\Delta= 1$AU. This magnitude given as 
\begin{equation}
m_{R}(1,1,0)=m_{R}-5log_{10} (r_n \Delta )- 2.5 log_{10} \Phi (\alpha)
\label{hglp}
\end{equation}
wher m$_{R}$ denotes apparent magnitude in R$_\mathrm{C}$-band. We adopted empirical  scattering phase function in order to correct the phase darkening effect;   $2.5 log_{10} \Phi (\alpha)=\beta \alpha$ , where   $\beta=0.035$  $(mag \; deg^{-1})$ is assumed.

\par An $Af\rho$ technique was used as proxy for the dust production rate \cite{ahearn1984cometbowell}. An $Af\rho$ is a product of band albedo $(A)$, filling factor $(f) $within aperture field of view and the linear radius of the aperture $\rho$, projected on the sky-plane at the object distance. 

\begin{equation}
Af\rho=4\frac{\Delta ^{2} r^{2}}{\rho} \frac{F_{obj}}{F_{\sun}}
\label{Afr}
\end{equation}
Here $F_{\sun} $ is the solar flux at 1 AU; $F_{obj}$ is the comet flux measured within the given aperture, $\Delta$ and $r $ are the geocentric and heliocentric distances, respectively.  The calculated magnitude, an $Af\rho$  parameter  and dust production rate are given in the Table \ref{pr:2}.

\begin{table}
\caption{Photometry Result: Comet 231P magnitude, $Af\rho$ parameter and dust production rate}
\label{pr:2}       
\begin{tabular}{c c c c c c c c}     
\hline 
\centering
UT date & $r_h$&$\Delta$ &$m_{R}$&$Af\rho$&$Q_{dust}$\\
& [AU]  & [AU] &[mag]&[cm] &[$kg s^{-1}$]\\
\hline \\
2014/07/25.85&4.88&4.02&$21.37\pm 0.08$&5.00&5 \\
\hline
\end{tabular}
\end{table}
\section{Discussion}
\label{dis}
This work based on the aperture photometric study and its derived parameter like dust production and cometary activity around aphelion. Previously study have not been studied in sufficient detail in this regard of QHCs. This is not good idea to compare our result with other member of QHCs, but it may comparable with JFCs. Please, see  the Table \ref{mafrho:3}  for more comparison.

\begin{table}
\caption{Measured  $Af\rho$ parameter and dust production rate}
\label{mafrho:3}       
\begin{tabular}{c c c c c c}     
\hline 
\centering
Comet & $r_h$&$Af\rho$&$Q_{dust}$&$References$ &\\
& [AU]  &[cm] &[$kg s^{-1}$]\\
\hline \\
231P&4.88&5.00&$5$& This work& \\
29P&6.25&3865.4&$431$& Shi J C  et al. (2014)& \\
29P&5.97&4637.4&$182$& Ivanova et al. (2009)& \\
228P&3.473&44&$6.9$& Shi J C  et al. (2014)& \\
228P&3.474&71.4&$11.2$& Shi J C  et al. (2014)& \\
117P&3.29&670&$-$& Epifani E M  et al. (2008)& \\
36P&3.90&11&$-$& Epifani E M  et al. (2007)& \\
129P&4.50&4.558&$-$&  Lowry \& Fitzsimmons (2005)& \\
74P&4.24&298&$-$&  Lowry \& Fitzsimmons (2001)& \\
159P&3.98&199&$-$& Epifani E M  et al. (2007)& \\
\hline
\end{tabular}
\end{table}
\par Since the main goal of our observing program was to search for signs of cometary activity in distance comets. We have calculated dust production rate using $Q_{dust}=\frac {Af\rho (4a_{dust} v_{ej}\sigma )}{3p} $ \cite{meech1996}and compared with available data of QHCs and JFCs; where $a_{dust}$ $(\mu m)$ is the grain radius, p is the geometric albedo of the dust grain, $v_{ej} (m s^{-1}) $ is the grain ejection velocity, $\sigma  $  $(g cm^{-3})$ is grain density. We assume, a= 1 cm (Grun 2001) \cite{grun2001broadband}, p=0.04,   $v_{ej}= 50$ $ m s^{-1}$ and $\sigma = 0.49  g cm^{-3} $ (Niimi 2012) \cite{niimi2011size}  but we adapted  $\sigma = 0.6  g cm^{-3} $ from Snodgrass C (2008) because of it is derived from JFCs and seem to be more realistic; the  dust production rate, $Q_{dust}=5  kg s^{-1} $,we found. As the dust production  rate calculated by Shi J (2014) \cite{shi2014ccd} for comet 228P/LINEAR on 2011, April 5 was $Q_{dust}=11.2  kg s^{-1} $, and on 2011, April 6 was $Q_{dust}=6.9  kg s^{-1} $, were as our value  $Q_{dust}= 5  kg s^{-1} $ is within the range as found by Shi J (2014), which is comparable with QHCs and seem to be more  realistic, under the above circumstance. 
\par The variation of the $Af\rho $  parameter and dust production rate mostly mean the variation of the activity of comets at different epoch. Observed comet 117P/Helin-Roman-Alu, $Af\rho=670 $ cm at 3.29 AU \cite{epifani2008distant}, similarly, comet 36P/Whipple, $Af\rho=11.0 $ cm at 3.9AU\cite{epifani2007distant}; comet 129P/Shoemaker-Levy-3 , $Af\rho=4.5 $ cm at 4.55 AU\cite{lowry2005william}; comet 74P/Smirnova-Chernykh , $Af\rho=298.0 $ cm at 4.24 AU\cite{lowry2001ccd} which appear to be large as we observed . 

\section{Concluding remarks}
\label{con}

The ground base observation program of Quasi-Hilda comet; 231P/LINEAR-NEAT was observed around aphelion passage, we attempted to address the question whether 231P active or not at aphelion. An apparent $R_{C}$-band magnitude was $21.37\pm 0.08 $ mag. The measured $Af\rho=5$ cm confirmed that the comet 231P was strongly dormant, according to the $Af\rho$ criterion. The measured dust production rate was about 5 $ kg s^{-1}$.
\section{Future Work}
\label{fwork}
 As a part of ground base observation program, we propose and planned to observe all member of QHC; in order to address it's cometary activity, dust production rate and mass lose rate around aphelion.

\section{acknowledgements}
The authors thank to prof. Masateru Ishiguro, SNU, South Korea to motivate and study Quasi-Hilda Comets, in order to address the cometary activity around aphelion. We also thank for  HCT observatory technical staff. 

\bibliographystyle{unsrt}       
\bibliography{231p}   
\end{document}